\documentclass{ws-ijmpa}
\usepackage{graphicx}
\usepackage{epsfig}

\begin{document}

\markboth{Philip Yock}
{Gamma-Gamma Interaction}

%
\catchline{}{}{}{}{}
%

\title{THE GAMMA-GAMMA INTERACTION : A CRITICAL TEST OF THE STANDARD MODEL}

\author{PHILIP YOCK}

\address{Department of Physics, University of Auckland, 38 Princess Street\\ Auckland, 1142, New Zealand\\
p.yock@auckland.ac.nz}

\maketitle

\begin{history}
\received{Day Month Year}
\revised{Day Month Year}
\end{history}

\begin{abstract}
Data from the Large Electron Positron collider (LEP) at CERN on hadron production in gamma-gamma interactions exceed the predictions of the standard model by an order of magnitude at the highest observed transverse momenta in three channels. The amplitude for the process is asymptotically proportional to the sum of the squares of the charges of quarks. The data are suggestive of models where quarks have unit charges, or larger, and where partons have substructure. A previously proposed model of electro-strong interactions includes both these features. Definitive measurements could be made with either of the linear electron-positron colliders that have been proposed, $viz.$ the International Linear Collider (ILC) or the Compact Linear Collider (CLIC). However, an electron-electron collider employing the recently developed ``plasma wakefield'' acceleration technique could provide the most affordable option. An independent check of the multi-muon events that were recently reported at Fermilab could also be made with this type of collider.

\end{abstract}

\ccode{PACS numbers: 11.10.Jj, 12.38.Qk, 13.66.Lm}

\section{Introduction}
Could the standard model of particle physics be fundamentally flawed? Despite its widespread use, the model  raises conceptual questions of naturalness, is not free of inconsistencies with observations, and includes unproven conjectures. The ever-present assumption that quarks can be treated simultaneously both as free and yet confined is particularly vexing. It distinguishes the standard model from the rest of science. Such assumptions are best tested in different situations. Similar tests are likely to yield similar results. 

Here, recently obtained data on high-energy $\gamma\gamma$ interactions are used to test the model. Hadron production in these interactions provides a new means for testing the model with attractive features. The initial state is simple, and, according to perturbation theory, the amplitude for the process depends on the squares of the charges of quarks. It thus depends sensitively on these fundamental parameters, and is also immune to possible cancellations between positive and negative charges. In addition, initial results from SLAC and DESY were mildly puzzling when compared to predictions of the standard model, thus rendering the process worthy of further investigation\cite{bin}. Recently, the process was studied at LEP with electron and positron beam energies of 100 GeV. These studies probed the highest currently available momentum transfers in $\gamma\gamma$ interactions, and the shortest distances. They are briefly described below.    	

\section{Gamma-Gamma Experiments at LEP}

The L3 collaboration reported results on $\gamma\gamma$ interactions in three channels. Data\cite{l3a} on inclusive $\pi^0$ production are shown in Fig. 1. They exceed the prediction of the standard model by about an order of magnitude at the highest transverse momenta. The $\gamma\gamma$ interactions occurred in the process $e^-e^+\:\rightarrow\:e^-e^+\pi^0X$ where $\gamma$'s were radiated by the incoming $e^-$ and $e^+$. The radiative process was known to be a copious source of $\gamma\gamma$ interactions\cite{bro}. 

\begin{figure}[pb]
\centerline{\psfig{file=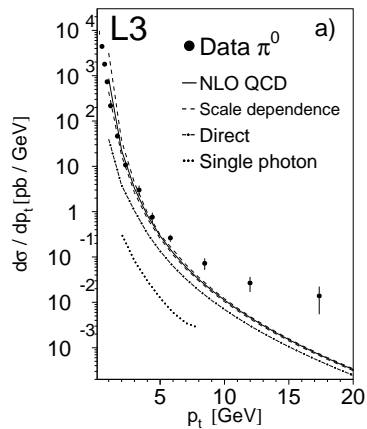,width=4.7cm}}
\vspace*{8pt}
\caption{Differerential cross-section for inclusive $\pi^{0}$ production in $\gamma\gamma$ interactions measured at LEP and the prediction of QCD (quantum chromodynamics). \label{f1}}
\end{figure}

Observations of $\pi^0$ production have not been reported by other groups. However, data on $\pi^{\pm}$ production have been reported by both L3\cite{l3b} and OPAL\cite{opa}. They are comparable to the $\pi^0$ data. Both L3 and OPAL reported significant excesses over the QCD prediction at $p_T\sim$ 17 GeV/c. The excess reported by L3 is slightly greater than that shown in Fig. 1 for $\pi^0$ production, whereas for OPAL the reported excess is slightly less. 

Data have also been reported by L3\cite{l3c} and OPAL\cite{opb} on inclusive jet production. The data by L3 extend to higher values of $p_T$ than those of OPAL, and they show a greater excess over the QCD prediction, reaching nearly an order of magnitude at 45 GeV/c.        

\subsection{Experimental Remarks}

The $\gamma\gamma$ data were derived from $e^-e^+$ interactions in which the incoming $e^-$ and $e^+$ radiated nearly real $\gamma$'s that interacted, and in which the outgoing $e^-$ and $e^+$ emerged in the beam pipe of the collider and went undetected. $\gamma\gamma$ events were identified through calorimetric measurements of the total energy of detected outgoing particles. This was significantly $<200$ GeV for $\gamma\gamma$ events.   

One source of background arose from $e^-e^+$ annihilation in which the final state energy was mismeasured. The main source of background arose, however, from $e^-e^+\;\rightarrow\;Z\gamma$ where initial state radiation reduced the invariant mass of the $e^-e^+$ system to that of the Z, and in which the $\gamma$ emerged in the beam pipe and was undetected. Hadronic decays of the Z in these events could mimic $\gamma\gamma$ events. However, Z's produced as above are boosted to an energy of $\sim130$ GeV. Both L3 and OPAL made cuts on the total energy measured in their calorimeters to remove this source of background. The high-precision calorimeters employed by L3 enabled them to place the cut at 80 GeV and maintain the background below $1\%$. In the case of OPAL, the cut had to be placed at a lower level, and/or extra conditions applied, resulting in analyses of greater complexity.                 

\section{Experimental Tests at Higher Energies}

The increasing nature of the above discrepancies between theory and observation is indicative of new substructure being uncovered in $\gamma\gamma$ interactions. From a purely observational point-of-view, and irrespective of any theoretical prejudice, the results call for testing at higher energies and higher momentum transfers.  There are at least two ways in which such tests could be accomplished.   

\subsection{Conventional Linear Colliders}

Two $e^-e^+$ linear colliders were proposed in recent years, the ILC and CLIC. The ILC was planned to reach collision energies of $0.5-1.0$ TeV. Although these would not greatly exceed the 0.2 TeV collison energy of LEP, the ILC would offer a distinct advantage. This would be the option to accelerate $e^-$'s in both arms of the collider, and thereby eliminate entirely the two main sources of background, described above, that occurred in the $\gamma\gamma$ experiments at LEP. In this way, momentum transfers up to the full kinematic limit of the collider could be reached. Unfortunately, however, funding opportunities for the ILC appear bleak\cite{bru}. The second linear collider that has been proposed, $viz.$ CLIC, is planned to reach collision energies from 0.2 to 5.0 TeV. CLIC could also be run in the $e^-e^-$ mode, but it is still at the research and development stage\cite{ell}.           

\subsection{Plasma Wakefield Collider}

An alternative procedure for reaching high collision energies in $\gamma\gamma$ interactions has emerged in recent years with encouraging features. This is the ``plasma wakefield'' technique in which intense electric fields are generated by a particle or laser pulse propagating through vapour, and used to accelerate charged  particles. The drive pulse ionises the vapour, leaving a positive ion beam along its track. The freed electrons are attracted back to the positive ion beam and, in so doing, generate a zig-zag wake with intense electric fields at the crossing points. These fields are used to accelerate particles. In a dramatic demonstration, 42 GeV electrons passing through lithium vapour were accelerated to energies up to 85 GeV in just 80-cm\cite{blu}. Two multistage electron accelerators based on this principal might generate $\gamma\gamma$ collisions at TeV energies for considerably less expense than the ILC and CLIC designs described above\cite{jos}.   

\begin{figure}[pb]
\centerline{\psfig{file=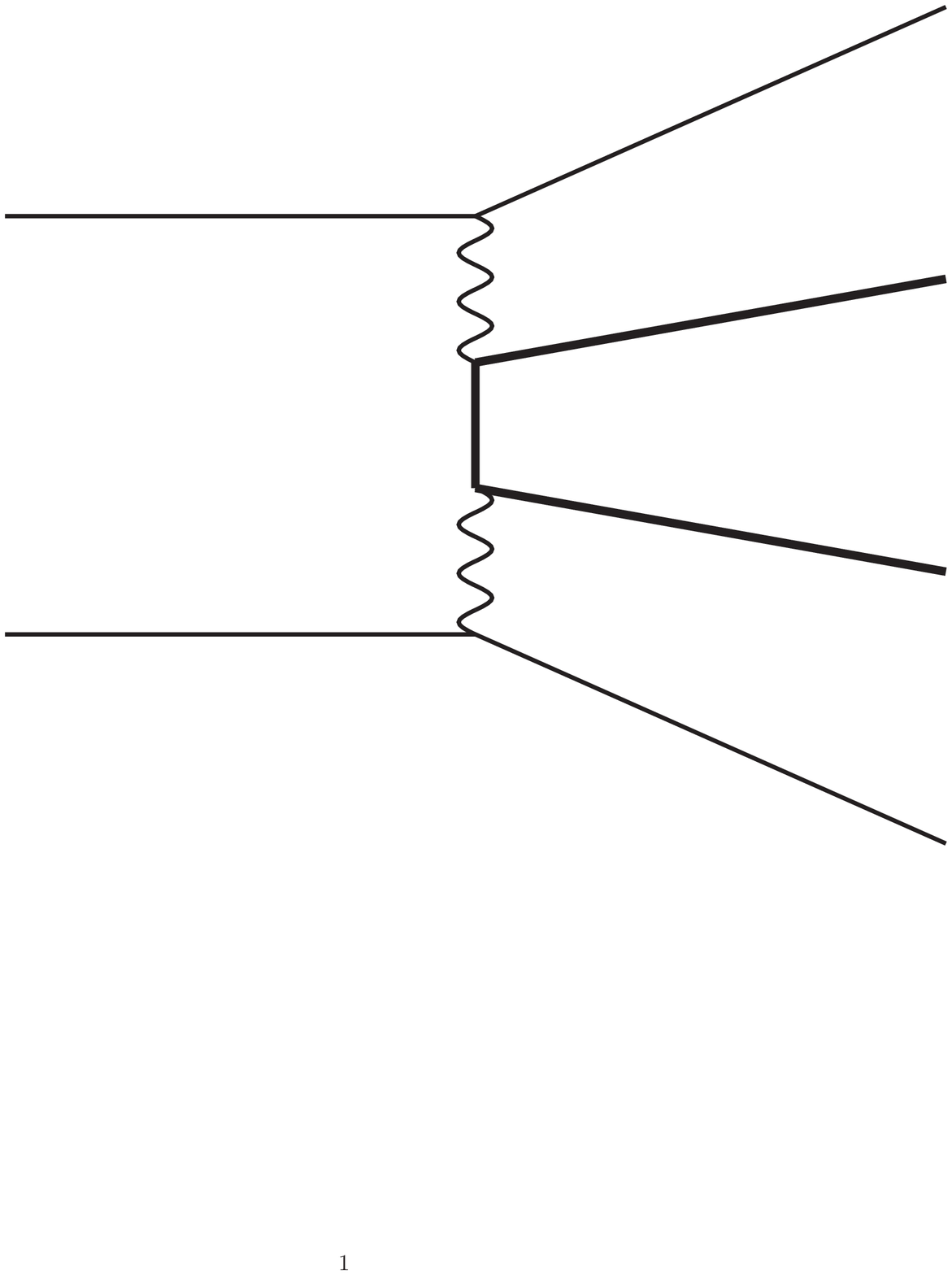,width=4.7cm}}
\vspace*{8pt}
\caption{Feynman diagram for hadron production at asymptotically high $p_T$ in $\gamma\gamma$ interactions as observed with $e^-e^+$ colliders. Thin lines denote leptons, heavy lines quarks. \label{f1}}
\end{figure}

\section{Physics Implications}

The $\gamma\gamma$ results from LEP are significant. According to the standard model, the Feynman diagram shown in Fig. 2 applies\cite{bin} when $p_T\geq$ few GeV/c. The amplitude for hadron production at large transverse momenta was therefore predicted to be proportional to the sum of the squares of the charges of quarks accessible at 200 GeV, and the cross section was expected to provide a simple but sensitive constraint on these basic charges\cite{wit}. The data, however, are inconsistent with this expectation, on two counts. Not only do the data exceed the predictions in all channels, but the excesses in each channel grow with transverse momentum. Quite generally, one might expect quarks with larger charges than the fractional values normally assumed to explain the first of these discrepancies, and the existence of substructure of some kind to explain the second.  

\subsection{Singly Charged Quarks}

Ferreira considered a Han-Nambu type of model with quarks having unit charges, and he found improved agreement with the data\cite{ferr}. However, the model did not include substructure of any form, and, as expected, the dependence of the data on transverse momentum was not well explained. Ferreira also provided a comprehensive review of the literature on quarks of unit charge.

\subsection{Highly Charged Quarks}

Before QCD was proposed, two theories of highly charged quarks were proposed, independently. In one, quarks were proposed by Schwinger\cite{scha} to be magnetic monopoles, in the other they were proposed\cite{yoca} to be highly electrically charged particles. In both cases, it was assumed that strong electromagnetic forces between quarks of opposite polarity accounted for quark binding, and that strong nuclear interactions between composite states occurred as residual interactions. Both theories were based on eigenvalue equations for charge. In Ref. 14 a modified version of Dirac's eigenvalue equation for magnetic charge was assumed. This predicts a large value for the electric dipole moment of the neutron, exceeding by several orders of magnitude the measured upper limit. In Ref. 15 a physical solution to a Gell-Mann-Low equation was assumed which required strong electromagnetic interactions, i.e. an ``electro-strong'' model. This is summarised below. 

\section{Electro-Strong Model}

\subsection{Formulation}

The guiding principle in the formulation of the electro-strong model was the assumption that the correct theory is finite without renormalization. In this regard we note that neither Feynman nor Dirac accepted the validity of conventional renormalization theory.  Conventional quantum electrodynamics, characterized by the Lagrangian   

\begin{equation}
L=e_{\circ}j_{\mu}^{e}A_{\mu},
\label{diseqn}
\end{equation}

\noindent may satisfy the finiteness criterion if the bare mass of the electron vanishes, and if the bare charge of the electron satisfies an eigenvalue equation of the form\cite{yoca} 

\begin{equation}
f(\alpha_{\circ})=-1.
\label{diseqn}
\end{equation}

\noindent Here $f(\alpha_{\circ})$ denotes the 4th + higher-order contributions to $Z_3^{-1}=e_{\circ}^2/e^2$, normalised by the 2nd order contribution, and $\alpha_{\circ}=e_{\circ}^2/4{\pi}^2$. A perturbative expansion of $f(\alpha_{\circ})$ yields

\begin{equation}
f(\alpha_{\circ})=O(\alpha_{\circ})+O({\alpha_{\circ}}^2)+O({\alpha_{\circ}}^3)+...
\label{diseqn}
\end{equation}

\noindent which is unlikely to be consistent with Eq. (2) if $\alpha_{\circ}\ll1$. 

In Ref. (16) a sequence of models was considered in an effort to avoid this impasse. The first was a model with strong interactions included following then-recent papers by Fermi and Yang\cite{fer}, Lee and Yang\cite{lee}, Sakurai\cite{sak}, Ne'eman\cite{nee} and Schwinger\cite{schb}. A vector field $B_{\mu}$ was introduced and coupled strongly to the proton current as below

\begin{equation}
L=g_{\circ}j_{\mu}^{p}B_{\mu}+e_{\circ}j_{\mu}^{p}A_{\mu}+e_{\circ}j_{\mu}^{e}A_{\mu},
\label{diseqn}
\end{equation}

\noindent where it was implicitily\footnote{We set $c=\hbar=1$.} assumed that $g_{\circ}^2/4{\pi^2}=\beta_{\circ}\sim O(1)$. With such a theory, Eq. (2) was found\cite{yoca} to bifurcate into a pair of conditions 

\begin{equation}
f(\beta_{\circ})\approx{-1}\:\:\rm{and}\:\: f(\beta_{\circ})\approx{-2}
\label{diseqn}
\end{equation}

\noindent that could not be simultaneously met. 

This motivated consideration of a unified gauge theory with the electromagnetic field $A_\mu$ coupled electromagnetically to the electron, and strongly coupled to a ``pseudo-proton'', as below  

\begin{equation}
L=g_{\circ}j_{\mu}^{p}A_{\mu}+e_{\circ}j_{\mu}^{e}A_{\mu}.
\label{diseqn}
\end{equation}

\noindent This resulted in a single condition for finiteness 

\begin{equation}
f(\beta_{\circ})\approx{-1}
\label{diseqn}
\end{equation}

\noindent which probably could be met, but an unrealistic theory in which the ``pseudo-proton'' was highly charged, i.e. $\beta_{\circ}\sim O(1)$. 

It was then but a small step to introduce the then-new quark concept by replacing the ``pseudo-proton'' field in Eq. (6) with a pair of quark fields, $Q$ and $q$, with bare charges $2g_{\circ}+e_{\circ}$ and $g_{\circ}$ respectively, as below

\begin{equation}
L=\{e_{\circ}j_{\mu}^{e}+(2g_{\circ}+e_{\circ})j_{\mu}^{Q}+g_{\circ}j_{\mu}^{q}\}A_{\mu}
\label{diseqn}
\end{equation}

\noindent where 

\begin{equation}
j_{\mu}^{e}=\overline{\psi}\gamma_{\mu}\psi,\;j_{\mu}^{Q}=\overline{Q}\gamma_{\mu}Q,\;j_{\mu}^{q}=\overline{q}\gamma_{\mu}q.
\label{diseqn}
\end{equation}

The theory defined by Eq. (8) includes both electromagnetic and strong interactions that are transmitted by the single vector field $A_{\mu}$. It this unifies strong and electromagnetic interactions, hence the terminology ``electro-strong'' model. Besides unifying strong and electromagnetic interactions, the model has other unusual features. These are discussed below. 

\subsection{Phenomenology}

An obvious feature of the electro-strong model is the inbuilt mechanism it possesses for quark binding, $viz.$, the strong attraction between highly charged quarks of opposite polarity that would be expected to form states that are neutral with respect to the high charge\cite{yoca}. This assumption of neutrality is analogous to the corresponding assumption that was made a few years later in QCD with respect to color charge\cite{fri}.   

\begin{figure}[pb]
\centerline{\psfig{file=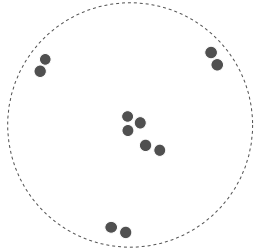,width=4.7cm}}
\vspace*{8pt}
\caption{Generalized Yukawa model of the nucleon. The dots represent highly charged quarks clustered to form bare mesons and nucleons. The overall scale is the $\pi$ Compton wavelength, i.e. the observed size of the nucleon. The smaller scale represents the fundamental separation between quarks in bound staes, which was assumed in Ref. 24 to be $\sim1000\times$ smaller. \label{f1}}
\end{figure}

The above mechanism for quark binding suggests the following particle classification for the model defined by Eq. (8):-    

\begin{eqnarray*}
e &=& \psi \\
\pi &=& \overline{Q}Q \:\: (l=0\; \rm{singlet\;state})\\
\rho &=& \overline{Q}Q \:\: (l=0\; \rm{triplet\; state})\\
\eta &=& \overline{q}q \:\: (l=0\; \rm{singlet\;state})\\
\phi &=& \overline{q}q \:\: (l=0\; \rm{triplet\; state})\\
p &=& Q\overline{q}\overline{q}
\end{eqnarray*}
and
\begin{equation}
\gamma=A_{\mu}.
\label{diseqn}
\end{equation} 

The spins and parities of the quark-antiquark states follow as in any fermion-antifermion model. The $\overline{q}q$ states would have different masses and interaction strengths from the $\overline{Q}Q$ states because of the different charges involved. Symmetry breaking in this sense is thus inbuilt in the model, without the introduction of a Higgs mechanism. Also, the presence of multiple generations (two in the above model) is required to form fermionic bound states such as the proton. This contrasts with the standard model where the presence of multiple generations is unexplained\cite{feyb}. For the proton state, antisymmetry of the $\overline{q}\overline{q}$ state requires this pair to be in the singlet spin state, which leads to a prediction of $J=1/2$ for the proton without the introduction of color. The proton's magnetic moment would be expected to be $\sim{g_0}/m_Q$ where $m_Q$ denotes the physical mass of the Q. For ${g_0}\gg{e}$ and ${m_Q}\gg{m_p}$ this ratio could be $\sim{e}/{m_p}$, suggesting that quarks may be heavy and tightly bound, as may have been originally supposed\cite{gela}. This suggests a small separation between quarks in bound states, and a generalized Yukawa model for hadrons\cite{yocb}, as indicated schematically in Fig. 3. 

Bare mesons and nucleons would act as partons\footnote{In this paper, ``quark'' is taken to signify fundamental constituents of hadrons, and ``parton'' the objects that cause interactions in deep-inelastic $ep$ processes at SLAC or similar energies} of unit charge in the generalized Yukawa model at non-asymptotic energies, without invoking the unexplained final state interactions of the standard model that allow fractionaly charged quarks to convert (or ``hadronize'') to integrally charged hadrons\cite{yocb}. However, the eventual onset of large electromagnetic effects at asymptotically high energies and high transervse momenta would be expected. These effects could make their first appearance in $\gamma\gamma$ interactions because these interactions are sensitive to the square of the charge of the quark, as discussed above. The growing discrepancy shown in Fig. 1 between the standard model and $\gamma\gamma$ data with increasing transverse momentum may be signaling the onset of these effects. Further studies of $\gamma\gamma$ interactions at higher energies and higher transverse momenta are called for.  

The apparent ease with which the electro-strong model appears potentially able to explain a number of phenomena, from divergences in quantum field theory, to quark statistics, quark binding, symmetry breaking, multiple generations, hadronization and, last but not least, the $\gamma\gamma$ data raises an obvious question, can the model reproduce the particle spectrum of the standard model? The answer, in the opinion of this author, is not positive\cite{yocc}. Although a qualitatively similar particle spectrum is possible, it is distinct. The difference arises from the different natures of strong interactions in the two models. In the electro-strong model, they are Abelian gauge interactions, in the standard model they are non-Abelian gauge interactions. The former require multiple generations of particles, and they include inbuilt symmetry breaking. The latter do not require multiple generations, and they require symmetry breaking to be added. Searches for the Higgs particle are therefore relevant.

Recent results from Fermilab on unexpected muons in $b\overline{b}$ events are also relevant\cite{mma}. The vertices of the unexpected muons are not coincident with $p\overline{p}$ interaction sites, indicating that the muons are decay products of long-lived, weakly-decaying particles. If so, the weakly-decaying particles apparently lie outside the particle spectrum of the standard model\cite{mmb}. We note here that $b\overline{b}$ production was also studied at CERN in $\gamma\gamma$ interactions. The results are comparable to those reported at Fermilab. The L3 group reported an excess of events over expectation which the ALEPH group attributed to secondaries not arising from the $\gamma\gamma$ interaction sites\cite{ale}. The similarity of the independently obtained results from Fermilab and CERN lends support to both datasets, and to the existence of particles lying outside the particle spectrum of the standard model. A new $\gamma\gamma$ collider as proposed above could test this possibility.       

\section{Concluding Remarks}  
Published data from CERN on $\gamma\gamma$ interactions at high energies are inconsistent with the current standard model of particle physics. The data exceed the predictions by about an order of magnitude in each of three channels. The discrepancies are greatest at the highest observed transverse momenta, and show no sign of decreasing, or even of levelling off, with increasing transverse momentum. The results question the fractional charges assumed for quarks in  the standard model, and suggest the existence of new substructure in hadrons that was not previously detected. Either the data are badly incorrect, or the standard model is seriously flawed. Irrespective of any theoretical prejudice, the data call for independent testing, preferably at higher energies and higher transverse momenta. Such testing could be carried with either of the linear colliders (ILC or CLIC) that have been proposed, but an electron-electron collider based on the plasma wakefield acceleration technique could provide the most affordable option.   

The $\gamma\gamma$ data may be compared to Rutherford's old data on $\alpha$-particle scattering by gold foil\cite{gei}. Although the physics underlying the $\alpha$-particle and $\gamma\gamma$ interactions is undoubtedly different, there may be a connection. Both cross-sections fall through several orders of magnitude with increasing transverse momentum, but in both cases the fall is gentler than was generally anticipated. It was, of course, Rutherford's experiment that supplanted J. J. Thomson's ``plumb-pudding'' model of the atom in which low-mass electrons swarmed in a cloud of massless, positive charge. It may be more than coincidental that the Thomson model bears resemblance to today's standard model of the nucleon in which low-mass quarks swarm in a cloud of massless colour charge\cite{shi}.  

The electro-strong model appears able to accommodate the $\gamma\gamma$ results, at least qualitatively. It provides motivation for the construction of a new, high-energy $\gamma\gamma$ collider of either conventional design, or using the plasma wakefield technique. This would follow the long-held viewpoint that the search for large electromagnetic effects at high energies provides the most reliable means for testing the electro-strong model\cite{yocc}.  

In recent years the author became involved in the search for habitable planets orbiting stars in the Galaxy\cite{moa}. This led naturally to thoughts on physics that might be underway on such ``exoplanets''. Others had similar thoughts. Gell-Mann expressed the opinion that the standard model would likely be under scrutiny\cite{gelb}. More recently, Rees\cite{ree} supported the case for strings. In the spirit\footnote{``The scientist does not study nature because it is useful; he studies it because he delights in it, and he delights in it because it is beautiful. If nature were not beautiful, it would not be worth knowing, and if nature were not worth knowing, life would not be worth living.'' \it{Henri Poincar$\acute{e}$}} of Poincar$\acute{\rm{e}}$, this author feels that simpler or more beautiful theories might yet be possible. Hopefully we need not wait to receive a blueprint from ``out there'' on the fundamental constituents of matter to decide the issue!

\section*{Acknowledgments}
Pablo Achard, Harald Fritzsch, Pedro Ferreira, Chan Joshi, Maria Kienzle-Focacci, Mark Kruse, Yvette Perrott, Francois Richard, Valery Telnov and Thorsten Wengler are thanked for discussion and correspondence. 

\section{References}


\end{document}